\documentstyle[a4,12pt]{article}
\input{psfig.sty}

\newcommand{\nc}{\newcommand}
\newcommand{\bQ}{\overline{Q}}
\newcommand{\ad}{\dot{\alpha }}
\newcommand{\bd}{\dot{\beta }}
\newcommand{\dd}{\dot{\delta }}
\nc{\beq}{\begin{equation}}
\nc{\eeq}{\end{equation}}
\nc{\bea}{\begin{eqnarray}}
\nc{\eea}{\end{eqnarray}}
\nc{\nn}{\nonumber}
\nc{\de}{\delta}
\nc{\bxi}{\bar{\xi}}
\nc{\bs}{\bar{\sigma}}
\nc{\bA}{\bar{A}}
\nc{\bpsi}{\bar{\psi}}
\nc{\bPsi}{\bar{\Psi}}
\nc{\bF}{\bar{F}}
\nc{\bl}{\bar{\lambda}}
\nc{\bW}{\bar{W}}
\nc{\bq}{\bar{q}}
\nc{\bel}{\bar{l}}
\nc{\bu}{\bar{u}}
\nc{\bh}{\bar{h}}
\nc{\bde}{\bar{d}}
\nc{\be}{\bar{e}}
\nc{\bchi}{\bar{\chi}}
\nc{\cL}{{\cal L}}
\nc{\cN}{{\cal N}}

\def\Dslash{D\!\!\!\! / \ }
\nc{\tr}{{\rm Tr}}
\nc{\halb}{\frac{1}{2}}

\def\cref{{\bf[add ref]}}

\begin{document}
\begin{titlepage}
\begin{center}

\hfill hep-th/9811341\\

\vskip .6in
{\large \bf The Supersymmetric Standard Model}
\vskip .5in

{\bf Jan Louis$^{a}$, Ilka Brunner$^{b}$ and 
Stephan J.\ Huber$^{c}$ }
\\
\vskip 0.8cm
{}$^{a}${\em Martin--Luther--Universit\"at 
Halle--Wittenberg,\\
        FB Physik, D-06099 Halle, Germany}\\
{\tt email:j.louis@physik.uni-halle.de}
\vskip 0.5cm
{}$^{b}${}{\em Department of Physics and Astronomy\\
Rutgers University\\
Piscataway, NJ 08855-0849, USA}\\
{\tt email:ibrunner@physics.rutgers.edu}
\vskip 0.5cm
{}$^{c}${\em Universit\"at Heidelberg,\\
     Institut f\"ur Theoretische Physik,\\
     Philosophenweg 16,\\
     D-69120 Heidelberg, Germany}\\
{\tt email:s.huber@thphys.uni-heidelberg.de}
\end{center}

\vskip 1.5cm

\begin{center} {\bf ABSTRACT } \end{center}
This set of lectures introduces
at an elementary level
the supersymmetric 
Standard Model and discusses some of its phenomenological
properties.\footnote{We intend to regularly
update this review. Therefore we would be very
grateful for your corrections, comments and suggestions.}

\vskip 2cm

\noindent
{\it Lectures given by J. Louis at the summer school
``Grundlagen und neue Methoden der 
theoretischen Physik'', Saalburg, 1996.}

\vfill
\noindent
November 1998
\end{titlepage}

\def\baselinestretch{1.2}
\baselineskip 16 pt

\section{Introduction}

The Standard Model of Particle Physics is an 
extremely successful
theory which has been tested  experimentally 
to a high level
of accuracy \cite{PDG,Lang}. 
After the discovery of the top quark 
the  Higgs boson which is predicted
to exist by the Standard Model
is the only `missing'
ingredient that has  not been directly 
observed yet.
 
However, a number of theoretical prejudices
suggest that the Standard Model is not the 
`final answer' of nature but rather
an effective description valid up
to the weak scale of order ${\cal O}(100 GeV)$.
The arbitrariness of the spectrum and gauge group, the 
large number of free parameters,
the smallness of the weak scale compared 
to the Planck scale and the
inability to turn on gravity 
suggest that at higher energies (shorter
distances) a more fundamental theory
will be necessary to describe nature.
Over the past 20 years 
various extensions  of the Standard Model
such as  Technicolor \cite{tech,FS}, Grand Unified Theories \cite{GGUT,ross}, 
Supersymmetry \cite{WZ,wb} or String Theory \cite{Green} have been proposed.
In recent years supersymmetric
extensions of the Standard Model became
very popular also among experimentalists
not necessarily because of their convincing solution of
the above problems but rather because 
most other contenders have been 
(more or less) ruled out by now.
Another reason for the popularity of 
supersymmetric theories among theorists
is the fact that the low energy
limit of superstring theory --
a promising 
candidate for a unification of all interactions including gravity 
-- is (by and large)
supersymmetric.

This set of lectures give an elementary introduction to the
supersymmetric Standard Model. Section~2 
contains some of the necessary background on 
generic supersymmetric field theories
while section~3 develops 
supersymmetric extensions of the Standard Model
and discusses spontaneous breaking of supersymmetry.
In section~4 extensions of the Standard Model 
with softly broken supersymmetry are presented
and some of the phenomenological 
properties are discussed. 
Section~5 contains a summary and 
our conventions which follow rather closely 
ref.\ \cite{wb} are recorded in an appendix. 

These lectures are not meant
to review the latest developments of the
supersymmetric Standard Model but rather
attempts to give an elementary introduction from 
a ``modern'' point of view.
Many excellent review articles on 
supersymmetry and the supersymmetric 
Standard Model do exist and have been heavily
used in these lectures \cite{RNi} -- \cite{Rsm}. 
In addition, a collection of 
some of the classic papers concerning the subject can be found 
in ref.~\cite{Fer}.

\section{Introduction to Supersymmetry}

Supersymmetry is a symmetry between bosons and fermions or more precisely it is a symmetry
between states of different spin \cite{WZ}.
For example, a spin-0 particle is mapped to 
a spin-$\frac{1}{2}$ particle 
under a supersymmetry transformation.
Thus,
the generators $Q_{\alpha }, \bQ_{\ad }$
of the supersymmetry transformation must 
transform in the 
spin-$\frac{1}{2}$ representations
 of the Lorentz group. 
These new fermionic 
generators form together with 
the four-momentum $P_m$
and  the generators of the Lorentz transformations $M^{mn}$
a graded Lie algebra
which features in addition to
commutators also anticommutators
in their defining relations.
The simplest ($N=1$)
supersymmetry algebra reads:
\bea \label{N1}
\{ Q_{\alpha},\bQ_{\bd} \} &=& 2\sigma_{\alpha \bd}^{m} ~P_m \nn\\ 
\{ Q_{\alpha} , Q_{\beta} \} 
&=& \{ \bQ_{\ad}, \bQ_{\bd} \} = 0\nn \\ 
{}[\bQ_{\ad}, P_m] &=& [Q_{\alpha} ,P_m ]=0\\ \nn
[Q_{\alpha}, M^{mn}] &=& \frac{1}{2}\,
\sigma^{mn \beta}_{\ \alpha} Q_\beta\\ \nn
[\bQ_{\ad}, M^{mn}] &=& \frac{1}{2}\,
\bar\sigma^{mn \dot\beta}_{\ \ad} Q_{\dot\beta}
\eea
where we used the notation and convention 
of ref.\ \cite{wb}.
 $\sigma^m$ are the Pauli matrices
and the $\sigma^{mn}$ are defined in the 
appendix.\footnote{
In general it is possible to have 
$N$ sets of supersymmetry generators
$Q_{\alpha }^I, \bQ_{\ad }^I, I=1,\ldots,N$,
in which case one refers to $N$-extended
supersymmetry. 
Such extended superalgebras have been
classified by 
Haag, Lopuszanski and Sohnius \cite{hls}
generalizing earlier work of 
Coleman and Mandula \cite{cm}
who showed that 
the possible bosonic symmetries of the 
S-matrix of a four-dimensional, local,
relativistic quantum field theory 
consist of the generators of the Poincar\'e
group and a finite number of generators 
of a compact Lie group,
which are Lorentz scalars.
In ref.\ \cite{hls} this theorem was generalized
to also include symmetry 
transformations generated
by fermionic operators and all possible
superalgebras were found. 
In extensions of the Standard Model
$N$-extended supersymmetries
have played no role  so far since
they cannot accommodate the chiral structure
of the Standard Model. }

The particle states in a supersymmetric field theory form representations (supermultiplets)
of the supersymmetry algebra (\ref{N1}).
We do not recall the entire 
representation theory here (see, for example, 
refs.\ \cite{RSo,wb}) 
but only highlight
a few generic features:

\begin{itemize} 
\item[(a)]
There is an equal number of bosonic degrees of freedom
$n_B$  and fermionic
degrees of freedom $n_F$
in a supermultiplet
\beq\label{nBF}
n_B = n_F\ .
\eeq

\item[(b)]
The masses of all states
in a supermultiplet are degenerate. 
In particular the masses
of bosons and fermions are 
equal\footnote{This follows immediately from the fact that 
$P^2$ is a Casimir operator of the supersymmetry
algebra (\ref{N1})
$[P^2,Q] = [P^2,M^{mn}]=0$.}
\beq\label{mass}
m_B = m_F\ .
\eeq

\item[(c)]
 $Q$ has mass dimension
$\frac{1}{2}$ and thus the mass dimensions
of the fields 
in a supermultiplet differ by $\frac{1}{2}$.

\end{itemize}

The two irreducible multiplets
which are important for constructing the supersymmetric Standard Model
are the chiral multiplet and the vector multiplet which we discuss in turn now.

\subsection{The chiral supermultiplet}

The chiral supermultiplet $\Phi$ \cite{WZ}
contains a complex scalar field $A(x)$ of spin 0 and mass dimension 1,
a Weyl fermion $\psi_{\alpha} (x)$ of spin $\frac{1}{2}$ and mass
dimension $\frac{3}{2}$ and an auxiliary complex scalar field $F(x)$ of spin 0 and 
mass dimension 2
\beq
\Phi = \big(A(x),\psi_{\alpha} (x),F(x)\big)\ .
\eeq
$\Phi$ has off-shell four real bosonic 
degrees of freedom ($n_B=4$)
and four real fermionic degrees of freedom
($n_F=4$) in accord with (\ref{nBF}).
The supersymmetry transformations
act on the fields in the multiplet as follows:
\bea \label{repr}
\de_{\xi} A &=& \sqrt{2} \xi \psi \nn  \\
\de_{\xi} \psi &=& \sqrt{2} \xi F 
     + i \sqrt{2} \sigma^m \bxi \partial_m A \\ \nn
\de_{\xi} F &=& i \sqrt{2} \bxi \bs^m \partial_m \psi
\eea
where we used the conventions of ref.\ \cite{wb}
and the appendix.
The parameters of the transformation
$\xi^{\alpha}$  are constant, complex
anticommuting Grassmann parameters obeying
\beq
\xi_{\alpha} \xi_{\beta}= - \xi_{\beta} \xi_{\alpha}.
\eeq
The transformations (\ref{repr})
can be thought of as 
generated by the operator
\beq\label{deldef}
\de_{\xi} = \xi Q + \bxi \bQ
\eeq
with $Q$ and $\bQ$ obeying (\ref{N1}).
This can be explicitly  checked by evaluating the commutators 
$[\de_{\xi}, \, \de_{\eta} ]$ on the fields
$A, \psi$ and $F$.\\

\noindent
{\it
{\bf Exercise:} 
Show $[ \de_{\xi}, \de_{\eta} ]
= 2i (\eta \sigma^m \bar\xi - \xi\sigma^m \bar\eta)
\partial_m $ by using (\ref{deldef}) and (\ref{N1}).\\

\noindent
{\bf Exercise:} Evaluate 
the commutator
$[ \de_{\xi}, \de_{\eta} ]$ using (\ref{repr}) 
for all three  fields $A, \psi$ and $F$
and show that this is  consistent 
with the results of the previous exercise.\\
}

The field $F$ 
has the highest mass dimension 
of the members of the chiral multiplet and therefore 
is called the highest component.
As a consequence it cannot transform 
into any other field
of the multiplet but only into
their derivatives.
This is not only true for the chiral
multiplet (as can be seen explicitly
in (\ref{repr})) but holds
for any supermultiplet.
This fact can be used to construct 
Lagrangian densities which transform into
a total derivative under supersymmetry
transformations leaving
the corresponding actions invariant.
We do not review here the method for systematically
constructing supersymmetric actions which 
is done most efficiently using
a superspace formalism.
Since these lectures focus on
the phenomenological properties of supersymmetry
we  refer the reader to the 
literature \cite{wb} for further details
and only quote the results.

For the chiral
multiplet a supersymmetric and  
 renormalizable Lagrangian 
is given by \cite{wb}
\bea
\cL (A,\psi,F) &=& 
-i \bpsi \bs^m \partial_m \psi - \partial_m \bA \partial^m A + F\bF  \nn\\
& &+m ( AF + \bA \bF -\frac{1}{2}(\psi \psi + \bpsi \bpsi) ) \\ \nn
& & +Y(A^2F + \bA^2 \bF - A\psi \psi - \bA \bpsi \bpsi)\ ,
\eea
where $m$ and $Y$ are real parameters.
This action has the peculiar property that
no kinetic term for $F$ appears.
As a consequence the equations of motion for $F$ are
purely algebraic 
$$
\frac{\de \cL}{\de \bF} = F +m\bA + Y \bA^2 = 0 ,
\qquad
\frac{\de \cL}{\de F} = \bF +mA + Y A^2 = 0.
$$
Thus 
$F$ is a non-dynamical, `auxiliary' field
which can be eliminated from the action
algebraically by using its 
equation of motion. This yields
\bea\label{Lafter}
\cL (A,\psi,F=-m\bA - Y\bA^2) &=& 
-i \bpsi \bs^m \partial_m \psi - \partial_m \bA \partial^m A 
 \\ \nn
&-&\frac{m}{2} (\psi \psi + \bpsi \bpsi)  -Y(A\psi \psi + \bA \bpsi \bpsi ) - V(A,\bA)
\eea
where $V(A,\bA)$ is the scalar
potential given by 
\bea\label{potential}
V(A,\bA) &=& \mid m A + Y A^2 \mid^2 \nn\\
&=& m^2A\bA + m Y(A\bA^2 + \bA A^2) + Y^2 A^2 \bA^2\\
&=& F\bF \mid_{\frac{\de \cL}{\de F}=\frac{\de \cL}{\de \bF}= 0} \ .\nn
\eea

As can be seen from (\ref{Lafter}) and
(\ref{potential}) after elimination of
$F$ a standard renormalizable Lagrangian
for a complex scalar $A$ and a Weyl fermion $\psi$
emerges. However (\ref{Lafter}) is not the most
general
renormalizable Lagrangian for such fields. 
Instead it satisfies the following
properties:
\begin{itemize}
\item
$\cL$ only depends on two independent
parameters, the mass parameter $m$
and the dimensionless Yukawa coupling $Y$.
In particular, the $(A\bA)^2$ coupling 
is not controlled by an independent
parameter 
(as it would be in non-supersymmetric theories)
but determined 
by the Yukawa coupling $Y$.

\item
The masses for $A$ and $\psi$ coincide 
in accord with (\ref{mass}).\footnote{As
immediate consequence of this feature
one notes that supersymmetry 
must be explicitly or
spontaneously broken 
in nature.}

\item
$V$ is positive semi-definite, $V \geq 0$.

\end{itemize}

\subsection{The vector supermultiplet}
The vector supermultiplet $V$ contains 
a gauge boson $v_m$ of spin 1 and
mass dimension 1,
a Weyl fermion (called the gaugino)
$\lambda $ of spin $\frac{1}{2}$
and mass dimension $\frac{3}{2}$,
and a real scalar field
$D$ of spin 0 
and mass dimension 2
\beq
V = (v_m(x), \lambda_{\alpha}(x), D(x))\ .
\eeq
Similar to the chiral multiplet
also the vector multiplet 
 has $n_B=n_F=4$.

The vector multiplet can be used to gauge the action of
the previous section.
An important consequence of the theorems 
of refs.~\cite{cm,hls} is the fact that the 
generators $T^a$ of a compact 
gauge group $G$ have to
commute with the supersymmetry generators
\beq
{}[T^a,Q_\alpha]=[T^a,\bQ_{\dot\alpha}]=0\ .
\eeq
Therefore all members of a chiral multiplet 
($A,\psi,F$) have to reside in the same
representation of the gauge group.
Similarly, the members of the vector  multiplet 
have to transform in the adjoint representation
of $G$ and thus they all are 
Lie-algebra valued  fields
\beq
v_m = v_m^a T^a\ , \qquad
\lambda_{\alpha } = \lambda_{\alpha}^a T^a \ ,
\qquad
D = D^a T^a \ .
\eeq

The
supersymmetry transformations of 
the components of the 
vector multiplet are \cite{wb}:
\bea \label{vsusy}
\de_{\xi} v_m^a &=& -i \bl^a \bs^m \xi + i \bxi \bs^m \lambda^a\ , \\ \nn
\de_{\xi} \lambda^a &=& i \xi D^a + \sigma^{mn}\xi F_{mn}^a\ , \\ \nn
\de_{\xi} D^a &=& - \xi \sigma^m D_m 
\bar\lambda^a - D_m \lambda^a \sigma^m \bxi\ .
\eea
The  field strength of the
vector bosons $F_{mn}^a$ 
and the covariant derivative $D_m \lambda^a$
are defined according to 
\bea
F_{mn}^a &:=& \partial_m v_n^a - \partial_n v_m^a - gf^{abc} v_m^b v_n^c\ ,\\
\nn
D_m \lambda^a &:=& \partial_m\lambda^a - gf^{abc} v_m^b \lambda^c\ ,
\eea
where $f^{abc}$ are the structure constants of the Lie algebra
and $g$ is the gauge coupling.
A gauge invariant, renormalizable
and supersymmetric
Lagrangian for the vector multiplet
is given by
\beq\label{puregauge}
\cL = -\frac{1}{4} F_{mn}^a F^{mn\,a} - i \bl^a \bs^m D_m \lambda^a 
+ \frac{1}{2} D^a D^a\ .
\eeq
As before
the equation of motion for the auxiliary
$D$-field is
purely algebraic $D^a = 0$.

A gauge invariant, renormalizable
Lagrangian containing a set of 
chiral multiplets ($A^i, \psi^i, F^i$)
coupled to vector multiplets
is found to be \cite{wb}
\bea \label{cvlagr}
\cL(A^i, \psi^i, F^i, v_m^a, \lambda^a, D^a)
&=& -\frac{1}{4} F_{mn}^a F^{mn ~a} - i\bl^a \bs^m D_m \lambda^a + \frac{1}{2}
 D^a D^a \nn \\ \nn
&& -D_m A^i D^m \bA^i - i \bpsi^i \bs^m D_m \psi^i
 + \bF^i F^i \\ 
&& +i \sqrt{2} g (\bA^i T_{ij}^a \psi^j \lambda^a 
  - \bl^a T_{ij}^a A^i \bpsi^j) \\ \nn
&& + gD^a\bA^i T_{ij}^a A^j - \frac{1}{2} W_{ij} \psi^i \psi^j 
   -\frac{1}{2} \bW_{ij} \bpsi^i \bpsi^j \\ \nn
&& +F^i W_i +  \bF^i \bW_i\ ,
\eea
where 
the covariant derivatives 
are defined by
\bea\label{covD}
D_m  A^i &:=& \partial_m A^i + ig v_m^a T_{ij}^a A^i\ , \\ \nn
D_m \psi^i &:=& \partial_m \psi^i + ig v_m^a T_{ij}^a \psi^j\ . 
\eea
$W_i$ and $W_{ij}$ in (\ref{cvlagr}) are
the derivatives of a holomorphic
function $W(A)$  called the superpotential
\bea \label{superpot}
W(A) &=& \frac{1}{2} m_{ij} A^i A^j + \frac{1}{3} Y_{ijk} A^i A^j A^k\ , \nn\\ 
W_i &\equiv& \frac{\partial W}{\partial A^i}= m_{ij} A^{j} +  Y_{ijk}A^j A^k\ ,  \\ \nn
W_{ij} &\equiv& \frac{\partial^2 W}{\partial A^i \partial A^j}
  = m_{ij} + 2 Y_{ijk} A^k\ .
\eea
By explicitly inserting (\ref{superpot})
into (\ref{cvlagr}) one observes that the
$m_{ij}$ are mass parameters while
the $Y_{ijk}$ are Yukawa couplings.
Supersymmetry forces $W$ to be a holomorphic 
function of the scalar fields $A$ while 
renormalizability restricts 
$W$ to be at most a cubic polynomial of $A$.
Finally, the parameters
$m_{ij}$ and  $Y_{ijk}$ are further constrained by
gauge invariance.

As before, $F^i$ and $D^a$ obey
algebraic equations of motion which read
\bea \label{FDeqm}
\frac{\de \cL}{\de F} = 0 & \Rightarrow & 
\bF_i + W_i = 0\ ,  \nn \\
\frac{\de \cL}{\de \bF} = 0 & \Rightarrow & F_i + \bW_i = 0\ , \\ \nn
\frac{\de \cL}{\de D^a} = 0 & \Rightarrow & D^a + g \bA^i T_{ij}^a A^j = 0\ .
\eea
They can be used to eliminate
the auxiliary fields $F^i$ and $D^a$ from 
the  Lagrangian (\ref{cvlagr}) 
and one obtains
\bea\label{cvlagrphys}
\lefteqn{
\cL(A^i, \psi^i, v_m^a, \lambda^a, F_i=-\bW_i, D^a= -g\bA^i T_{ij}^a A^j)=} \nn\\
&&-\frac{1}{4} F_{mn}^a F^{mn ~a} - i\bl^a \bs^m D_m \lambda^a  
  -D_m A^i D^m \bA^i - i \bpsi^i \bs^m D_m \psi^i  \\ \nn
&& +i \sqrt{2} g (\bA^i T_{ij}^a \psi^j \lambda^a 
  - \bl^a T_{ij}^a A^i \bpsi^j) 
 - \frac{1}{2} W_{ij} \psi^i \psi^j 
   -\frac{1}{2} \bW_{ij} \bpsi^i \bpsi^j - V(A,\bA)
\eea
where
\bea \label{scalarpot} 
V(A,\bA) &=& W_i \bW_i 
+ \frac{1}{2} g^2(\bA^i T^a_{ij}A^j)(\bA^i T^a_{ij}A^j)  \nn\\
         &=& (F^i \bF^i + \frac{1}{2} D^a D^a)\mid_{\frac{\de \cL}{\de F}=0,
                                                   \frac{\de \cL}{\de D^a}=0}\\
         &\geq & 0 \ .\nn
\eea
As before the scalar potential
$V(A,\bA)$ is positive semi-definite.

\bigskip
\noindent
{\bf Exercise:} {\it Insert
(\ref{FDeqm}) into (\ref{cvlagr}) and derive (\ref{cvlagrphys}).}

\section{A Supersymmetric Extension of the 
Standard Model}
\subsection{The  Standard Model}

In this section we briefly review
some basic features of the Standard Model.
The Standard Model is a 
quantum gauge field 
theory with a  chiral gauge group  
$G_{\rm SM}=SU(3)\times SU(2)\times U(1)_Y$.
The spectrum of particles includes 
three families of quarks and leptons,
the gauge bosons (gluons, $W^\pm, Z^0$, photon)
of $G_{\rm SM}$
and one spin-0 Higgs doublet.
In table~1 the particle content and the
corresponding gauge quantum numbers are displayed.
\begin{table} 
\begin{center}
\vspace{0.4cm}
\begin{tabular}{|l||c||ccc|c|} 
\hline
&& SU(3) & SU(2) & U(1)$_Y$ & U(1)$_{\rm em}$ \\[0.5ex]
\hline
&&&&& \\
quarks 
& $q_L^I = \left( \begin{array}{c} u_L^I \\ d_L^I \end{array} \right)$&
3 & 2 & $\frac{1}{6}$ & $\left( \begin{array}{c} \frac{2}{3} \\ -\frac{1}{3}
                      \end{array} \right) $ \\[0.5ex]
&&&&& \\
& ${u}_R^I $&  
$\bar{3}$ & 1 & $-\frac{2}{3}$ & $-\frac{2}{3}$\\[0.5ex]
& ${d}_R^I$ &
$\bar{3}$ & 1 & $\frac{1}{3} $ & $\frac{1}{3}$ \\[1ex]
&&&&&\\
\hline
&&&&&\\
leptons &
$l_L^I = \left( \begin{array}{c} \nu_L^I \\ e_L^I \end{array} \right)$ &
1 & 2 & $ -\frac{1}{2} $& $\left( \begin{array}{c} 0 \\ -1 \end{array}
\right)$ \\[0.5ex]
&&&&& \\
& ${e}_R^I$ &
1 & 1 & $1$ & $1$ \\[1ex]
&&&&&\\
\hline
&&&&&\\
Higgs &
$ h= \left( \begin{array}{c} h^0 \\ h^- \end{array} \right)$ &
1 & $2$ & $-\frac{1}{2} $ & $\left( \begin{array}{c} 0 \\ -1 \end{array} 
\right)$ \\
&&&&& \\
\hline
&&&&& \\
gauge bosons& $G$& 8&1&0&0\\
&$W$ &1&3&0&$(0,\pm1)$\\
&$B$&1&1&0&0\\
&&&&& \\
\hline
\end{tabular}
\caption{The particle content of the 
Standard Model. 
The index $I=1,2,3$  labels the three families
of chiral quarks $q_L^I, {u}_R^I, 
{d}_R^I$  and chiral leptons $l_L^I, {e}_R^I$. 
All of them are Weyl fermions and transform in the 
$(\halb,0)$ representation of the Lorentz group
(they have an undotted spinor
index $\alpha$). 
The subscripts $R,L$ do not specify the representation
of the Lorentz group but instead
 are used to indicate
the different transformation properties
under the chiral
gauge group $SU(2)\times U(1)$.  
This somewhat unconventional notation is used
to make a smooth transition to the supersymmetric
Standard Model later on.
The electromagnetic charge listed in the last column is defined by
$Q_{em} = T^3_{SU(2)} + Q_Y $.}
\end{center}
\end{table}
The  Lagrangian of the Standard Model reads
\bea\label{LSM}
\cL &=& -\frac{1}{4} \sum_{(a)=1}^{3} 
           \left( (F_{mn}^b F^{mn ~b})_{(a)} \right) - D_m h D^m \bh \nn\\
&& + \sum_{I=1}^3 \left( -i \bar{q}_L^I \Dslash q_L^I -i  \bu_R^I \Dslash u_R^I
   - i\bar{d}_R^I \Dslash d_R^I 
-i \bel_L^I \Dslash l_L^I -i \be_R^I \Dslash e_R^I \right) \\ \nn
&& - \sum_{IJ=1}^3\left( ({Y_u})_{IJ} \bh q_L^I u_R^J 
+ ({Y_d} )_{IJ} h q_L^I d_R^J
   + ({Y_l})_{IJ} h l_L^I e_R^J + h.c.\right)  -V(h,\bh),
\eea
where $\Dslash= \sigma^m D_m $
and the index $(a)$ labels the 3 different
factors in the gauge group. $V(h,\bh)$ is 
the scalar potential  for the
Higgs doublet which is chosen to be 
\beq
V(h,\bh)= \mu^2 h \bh + \lambda (h \bh)^2\ . 
\eeq
In order to have a bounded
potential
$\lambda>0$ has to hold. For $\mu^2 <0$ the electroweak
gauge group 
$SU(2)\times U(1)_Y$ is spontaneously
broken down to $U(1)_{\rm em}$.
In this case the minimum of the potential is not at $\langle h\rangle =0$, but at
$\langle h \bar{h}\rangle 
= -\frac{\mu^2}{2\lambda}.$

\bigskip

\noindent
{\bf Exercise:} {\it Give explicitly
all covariant derivatives in (\ref{LSM}).}

\noindent
{\bf Exercise:} {\it Check that
the Lagrangian  (\ref{LSM})
is gauge and Lorentz invariant.}

\subsection{Supersymmetric Extensions} \label{ssm}
Let us now turn to the supersymmetric generalization
of the Standard Model.\footnote{See also
\cite{RNi} - \cite{Fer}.}
The idea is to promote the Lagrangian (\ref{LSM})
to a supersymmetric Lagrangian.
As we learned in the previous section
supersymmetry requires the presence of additional
states which form supermultiplets with
the known particles. Since all states 
of a supermultiplet carry 
the same gauge quantum numbers we need
at least a doubling of states:
For every field of the SM one has to 
postulate a superpartner 
with the exact same gauge quantum numbers 
and a spin such that it can form an
appropriate supermultiplet.
More specifically, the quarks and leptons
are promoted to chiral multiplets
by adding scalar (spin-0) squarks 
($\tilde{q}_L^I, \tilde{u}_R^I, \tilde{d}_R^I$)
and sleptons ($\tilde{l}_L^I, \tilde{e}_R^I$)
to the spectrum.
The gauge bosons are promoted to vector multiplets by adding the corresponding spin-$\halb$
gauginos 
($\tilde{G}, \tilde{W}, \tilde{B}$) to the spectrum.
Finally, the Higgs boson is also
promoted to a chiral multiplet
with a spin-$\halb$ Higgsino superpartner.
However, the supersymmetric version of the 
Standard Model cannot `live' with only 
one Higgs doublet and at least a
second Higgs doublet has to be added.
This can be seen from the fact
that  one cannot write down 
a supersymmetric version of the 
Yukawa interactions of the Standard Model
without introducing a second Higgs doublet. 
The reason is 
the definite chirality of the Higgsino.
Another way to see the necessity
of a second Higgs doublet is the fact that
 the Higgsino
is a chiral fermion which carries $U(1)$ hypercharge and 
hence it upsets the anomaly cancellation
condition.
Thus a second Higgsino with opposite $U(1)$
charge
is necessary and supersymmetry 
then also requires a second
spin-0 Higgs doublet.\footnote{%
 Of course, extensions with more
Higgs doublets are  also possible, but 
two is the  minimal number.} 
The precise spectrum of the supersymmetric
Standard Model is summarized in
table~2.

\begin{table} 
\begin{center}
\begin{tabular}{|l||c||cc||ccc|c|} 
\hline
&supermultiplet& F & B & SU(3) & SU(2) & U(1)$_Y$ & U(1)$_{\rm em}$ \\
\hline
&&&&&&& \\
quarks 
& $Q_L^I = \left( \begin{array}{c} U_L^I \\ D_L^I \end{array} \right)$&
$q_L^I$ & $ \tilde{q}_L^I $ &
3 & 2 & $\frac{1}{6}$ & $\left( \begin{array}{c} \frac{2}{3} \\ -\frac{1}{3}
                      \end{array} \right) $ \\
&&&&&&& \\
& ${U}_R^I $&
$ {u}_R^I $ & $ \tilde{{u}}_R^I $ &  
$\bar{3}$ & 1 & $-\frac{2}{3}$ & $-\frac{2}{3}$\\[0.5ex]
& ${D}_R^I$ & $ {d}_R^I $ & $ \tilde{{d}}_R^I $ &
$\bar{3}$ & 1 & $\frac{1}{3} $ & $\frac{1}{3}$ \\[1ex]
&&&&&&&\\
\hline
&&&&&&&\\
leptons &
$L_L^I = \left( \begin{array}{c} \cN_L^I \\ E_L^I \end{array} \right)$ &
$l_L^I $ & $  \tilde{l}_L^I $ &
1 & 2 & $ -\frac{1}{2} $& $\left( \begin{array}{c} 0 \\ -1 \end{array}
\right)$ \\
&&&&&&& \\
& ${E}_R^I$ &
$ {e}_R^I $ & $ \tilde{{e}}_R^I $ &
1 & 1 & $1$ & $1$ \\[1ex]
&&&&&&&\\
\hline
&&&&&&&\\
Higgs &
$ H_d= \left( \begin{array}{c} H_d^0 \\ H_d^- \end{array} \right)$ &
$  \left( \begin{array}{c} \tilde{h}^0 \\ \tilde{h}^- \end{array} \right)$ &
$  \left( \begin{array}{c} h_d^0 \\ h_d^- \end{array} \right)$ &
1 & 2 & $-\frac{1}{2} $ & $\left( \begin{array}{c} 0 \\ -1 \end{array} 
\right)$ \\
& $ H_u= \left( \begin{array}{c} H_u^+ \\ H_u^0 \end{array} \right)$ &
$  \left( \begin{array}{c} \tilde{h}^+ \\ \tilde{h}^0 \end{array} \right)$ &
$  \left( \begin{array}{c} h_u^+ \\ h_u^0 \end{array} \right)$ &
1 & 2 & $\frac{1}{2} $ & $\left( \begin{array}{c} 1 \\ 0 \end{array} 
\right)$ \\
&&&&&&& \\
\hline
&&&&&&& \\
gauge& $G$&$\tilde G$&$G$ & 8&1&0&0\\
bosons&$W$&$\tilde W$ &$W$&1&3&0&$(0,\pm1)$\\
&$B$&$\tilde B$&$B$&1&1&0&0\\
&&&&&&& \\
\hline
\end{tabular}
\caption{Particle content of the supersymmetric Standard Model. The column below `F' (`B')
denotes the fermionic (bosonic)
content of the model.}
\end{center}
\end{table}

The Lagrangian for the supersymmetric Standard
Model has to be of the form (\ref{cvlagr})
with an appropriate superpotential $W$.
It has to be chosen such that the 
Lagrangian of the non-supersymmetric
Standard Model (\ref{LSM}) is contained.
This is achieved by 
\beq \label{mssmpot}
W= \sum_{IJ}\left(({Y_u})_{IJ} h_u \tilde{q}_L^I \tilde{u}_R^J + 
   ({Y_d})_{IJ} h_d \tilde{q}_L^I \tilde{d}_R^J +
   ({Y_l})_{IJ} h_d \tilde{l}_L^I \tilde{l}_R^I
\right) + \mu\, h_u h_d\ .
\eeq

Once $W$ it specified also the scalar potential
is fixed. Of particular importance is the
scalar potential for the Higgs
fields since it controls the electroweak
symmetry breaking.
Using (\ref{scalarpot}) and (\ref{mssmpot})
one derives the Higgs potential for the
two neutral Higgs fields $h_d^0, h_u^0$
by setting all other scalars to zero\footnote{Note that the scalars
can only be set to zero in the potential $V$
but not in the superpotential $W$ since the computation of
the potential requires taking appropriate derivatives of $W$.}
\beq\label{HiggsV}
V(h_d^0, h_u^0) =\ |\mu|^2 \left(|h_d^0|^2 + |h_u^0|^2 \right)
  +\frac{1}{8} \left( g_1^2 + g_2^2 \right) 
  \left( |h_u^0|^2 - |h_d^0|^2 \right)^2\ .
\eeq
The coupling
of the terms quartic in the Higgs fields 
is not an independent parameter but instead  
determined by
the gauge couplings $g_1$ of $U(1)_Y$
and $g_2$ of $SU(2)$.
Thus it seems that the number of parameters
is reduced.
However, now there are two possible
vacuum expectation values 
$\langle h_u^0\rangle, \langle h_d^0\rangle$
-- one more than in the Standard Model.

\bigskip

\noindent
{\bf Exercise:} {\it Derive (\ref{HiggsV})
from (\ref{scalarpot}) and (\ref{mssmpot}).}

\bigskip

In the last section we learned that 
the potential of any supersymmetric theory
is positive semi-definite
and the Higgs potential of eq.~(\ref{HiggsV})
is no exception as can be seen explicitly:
$|\mu|^2$ cannot be
chosen negative.
Thus the minimum 
of $V$ necessarily sits at 
$\langle h_u^0 \rangle =
\langle h_d^0 \rangle =0$
which corresponds to a vacuum with
unbroken $SU(2)\times U(1)$.
Therefore, the supersymmetric version
of the Standard Model as it is defined so far
-- the spectrum of table~2 with interactions
specified by the Lagrangian
 (\ref{cvlagr}) with the  $W$ of
(\ref{mssmpot}) --
cannot accommodate a vacuum with 
spontaneously broken electroweak symmetry.
A second phenomenological problem is the
presence of all the new supersymmetric
states which have the same mass as their 
superpartners but are not observed in nature.
As we said before, supersymmetry itself
necessarily has to
 appear in its broken phase
and as we will see electroweak symmetry breaking
is closely tied to the breakdown of supersymmetry.

Before we close this section let us note 
that in addition to the couplings
of (\ref{mssmpot})
gauge and Lorentz invariance
 also allows terms in $W$ which are 
of the form
\beq \label{nono}
h_u \tilde{l}_L,\quad  \tilde{l}_L \tilde{q}_L \tilde{d}_R, \quad
\tilde{d}_R \tilde{d}_R \tilde{u}_R,\quad
 \tilde{l}_L \tilde{l}_L \tilde{e}_R\ .
\eeq
These terms violate baryon or lepton number conservation and thus easily lead
to unacceptable physical consequences
(for example the proton could become unstable
\cite{Wein}). 
Such couplings can be excluded by imposing
a  discrete R-parity \cite{Fa}. 
Particles of the Standard
Model (includuing both Higgs doublets)
are assigned  R-charge 1 while
all new supersymmetric particles 
are assigned R-charge $-1$.
This eliminates all terms of (\ref{nono})
while the superpotential of (\ref{mssmpot})
is left invariant.
An immediate consequence of this additional
symmetry is the fact that
the lightest supersymmetric
particle (often denoted  by the `LSP')
is necessarily stable.
However, one should stress that R-parity is not
a phenomenological necessity.
Viable models with broken R-parity can be 
constructed and they also can have some 
phenomenological appeal \cite{brokenR}.

\bigskip

\noindent
{\bf Exercise:}
{\it Check the gauge and Lorentz invariance
for each term in (\ref{nono})
and compute their R-charge.}

\bigskip

\subsection{Spontaneous breaking of supersymmetry} \label{spon}

In the previous section we learned that 
in the simplest supersymmetric extension
of the Standard Model the electroweak symmetry is
unbroken.
However, so far we constructed a manifestly
supersymmetric extension but from the mass degeneracy
of each multiplet (\ref{mass})
it is already clear that supersymmetry
cannot be an exact symmetry in nature
but has to be either
spontaneously or explicitly broken.
Therefore we now turn to the question
of spontaneous supersymmetry breaking
and return to the electroweak symmetry breaking 
afterwards.

Let us first recall the order parameter
for supersymmetry breaking.
Multiplying the anticommutator
$
\{ Q_{\alpha}, \bQ_{\ad} \} = 2 \sigma_{\alpha \ad}^m P_m
$ of the supersymmetry-algebra (\ref{N1})
with $\bs^n$ and using 
$ Tr(\sigma^m \bs^n) = -2\eta^{mn}$
results in 
$$
\bs^{n \alpha \ad } \{ Q_{\alpha}, \bQ_{\ad} \} = -4P^n\ .
$$
Thus the Hamiltonian $H$ of a supersymmetric
theory is expressed as the  `square'
of the supercharges
\beq \label{ham}
H= P_0 = \frac{1}{4} \left( Q_1 \bQ_1 + \bQ_1 Q_1 + Q_2 \bQ_2 + \bQ_2 Q_2
\right)\ .
\eeq
This implies that
$H$ is a positive semi-definite operator
on the Hilbert space 
\beq
\langle \psi |H|\psi \rangle \; \geq 0 \; 
\quad \forall \psi\ . 
\eeq
Supersymmetry is unbroken
if the supercharges annihilate the vacuum
$Q_\alpha|0\rangle = \bQ_{\dot\alpha}
|0\rangle = 0$. From (\ref{ham}) we learn
that also $H$ annihilates a
supersymmetric  vacuum $H|0\rangle = 0$.
This in turn implies that 
the scalar potential $V$ of 
a supersymmetric field theory
which has a supersymmetric ground state
has to vanish at its minimum
\beq\label{Vmin}
\langle H\rangle = 0\quad \Rightarrow\quad 
\langle  V \rangle \equiv
V(A, \bA) |_{\rm min} = 0\ .
\eeq
The general form of the scalar potential 
$V=  F^i \bF^i + \frac{1}{2} D^a D^a$
was given in (\ref{scalarpot}). 
Since $V$ is 
positive semi-definite one immediately
concludes from (\ref{Vmin}) 
that in a supersymmetric ground state
\beq
\langle F^i\rangle \equiv F^i |_{\rm min} = 0 \quad {\rm and} \quad 
\langle  D^a \rangle \equiv D^a |_{\rm min} = 0
\eeq
has to hold.
The converse is also true
\beq
\langle F^i\rangle \neq 0 \quad {\rm or}\quad
\langle  D^a \rangle \neq 0\quad
\Rightarrow\quad
V|_{min} >0\quad \Rightarrow\quad
 Q_{\alpha} |0>\ \neq 0\ 
\eeq
and supersymmetry is spontaneously broken.
Thus 
$\langle F^i\rangle$ and $\langle D^a\rangle$ 
are the order parameters of supersymmetry breaking
in that non-vanishing $F$- or $D$-terms signal
spontaneous supersymmetry breaking.

Specific potentials which do lead 
to non-vanishing $D$-  or $F$-terms
have been constructed \cite{FIl,ORa}.
For example,
the  O'Raifeartaigh model \cite{ORa} has
three chiral superfields $A_0, A_1, A_2$
and the following superpotential:
\beq\label{WOR}
W = \lambda A_0 + m A_1 A_2 + g A_0 A_1^2\ , \qquad  m^2>2\lambda g\ .
\eeq
By minimizing $V$ it can be shown that 
$F_0|_{min} \neq 0$ and therefore supersymmetry is broken.
Furthermore the mass spectrum of the
6 real bosons and the 3 Weyl fermions is
found to be
\bea\label{mspectrum}
{\rm Bosons:} \quad && (0,0,m^2,m^2, m^2 \pm 2\lambda g) \\ \nn
{\rm Fermions:} \quad && (0,m^2,m^2)\ .
\eea
Thus the mass degeneracy between bosons
and fermions is lifted but nevertheless
a `mass sum rule' still holds
\beq \label{bosferm}
\sum_{\rm bosons} M_{b}^2 
= 2 \sum_{\rm fermions} M_{f}^2\ .
\eeq

\bigskip

\noindent
{\bf Exercise:} 
{\it Minimize $V$ using
(\ref{scalarpot}), (\ref{WOR}) 
and compute $F_i|_{\rm min}$.
 Verify the mass spectrum (\ref{mspectrum})
and the sum rule (\ref{bosferm}).}

\bigskip

Unfortunately,  the sum rule 
(\ref{bosferm}) is not a coincidence but 
a special case of a general sum rule
which
holds in any theory with spontaneously broken
supersymmetry. 
Let us therefore proceed and derive 
this sum rule.
In general the mass matrix of the bosons
has the following form
\bea
V(A,\bA )|_{\rm mass\ terms}\ =\  \mu^2_{i\bar{j}} 
A^i \bA^j + \mu^2_{ij} A^i A^j 
+ \mu^2_{\bar{i} \bar{j}} \bA^i \bA^j 
\ =\ ( \bA \quad A ) M_0^2
   \left( \begin{array}{c} A \\ \bA \end{array} \right) 
\eea
where
\beq\label{mmatrix}
M_0^2 = \left( \begin{array}{c c} \frac{1}{2} \mu^2_{i\bar{j}} &
                                 \mu^2_{ij} \\ 
                                 \mu^2_{\bar{i} \bar{j}} &
                                 \frac{1}{2} \mu^2_{i\bar{j}}
                \end{array} \right)\ .
\eeq
The entries in the mass matrix are 
determined by 
the appropriate derivatives of the potential 
evaluated at its minimum
\beq
\mu^2_{i\bar{j}} = V_{i\bar{j}}|_{min}\ , \qquad
\mu^2_{ij} = V_{ij}|_{min}\ , \qquad
\mu^2_{\bar{i} \bar{j}} = V_{\bar{i} \bar{j} }|_{min}\ ,
\eeq
where $V_{i\bar{j}}\equiv {\partial^2 V\over
\partial A^i \partial \bar{A}^{\bar{j}}}$ etc.
Using (\ref{scalarpot}) one derives
\bea
V &=& W_i \bW_{i} + \frac{1}{2} D^a D^a\ , \nn\\
V_j &=& W_{ij}  \bW_{i} + D^a_j D^a\ , \\ \nn
V_{j \bar{k}} &=& W_{ij} \bW_{ik} + D^a_j D^a_{\bar{k}} 
 + D^a D^a_{j \bar{k}}\ ,
\eea
where again the indices $i,j,\ldots$
denote derivatives with respect to $A^i, A^j, \ldots$.
Inserted into (\ref{mmatrix})
one obtains for the trace of the mass matrix:
\beq\label{bosonictrace}
\tr M_0^2 = \tr \mu^2_{i\bar{j}} 
=\tr V_{i \bar{j} }|_{min}
= \tr(W_{ij} \bW_{{i}{k}} 
+ D^a_j D^a_{\bar{k}} 
+ D^a D^a_{j \bar{k}})|_{min}\ .
\eeq

For the fermion masses the relevant pieces of the Lagrangian (\ref{cvlagr}) are
\beq
\cL = i \sqrt{2} g \; (\, \bA^i T_{ij}^a \psi^j \lambda^a 
  - \bl^a T_{ij}^a A^i \bpsi^j \,) 
- \frac{1}{2} W_{ij} \psi^i \psi^j 
   -\frac{1}{2} \bW_{ij} \bpsi^i \bpsi^j + \ldots\ .
\eeq
This can be rewritten as
\beq
\cL = 
-\frac{1}{2} ( \psi^i\ \lambda^a) M_{1/2} 
 \left(\begin{array}{c} \psi^j \\ \lambda^b \end{array} \right)  
\ +  {\rm h.c.} \ + \ldots
\eeq
where
\beq \label{fermmass}
M_{1/2} = \left( \begin{array}{cc}
W_{ij} & -i\sqrt{2} g \bA^i T_{ij}^{a} \\
-i\sqrt{2} g \bA^i T_{ij}^b & 0 \end{array} \right)
= \left( \begin{array}{cc}
W_{ij} & i\sqrt{2} D_j^{a} \\
i\sqrt{2} D_i^b & 0 \end{array} \right)\ .
\eeq
Thus, we obtain 
\beq\label{fermionictrace}
\tr M_{1/2} \bar{M}_{1/2}
= \tr (W_{ik} \bW_{{k}{j}} 
+ 4 D_i^a D_{\bar{j}}^a )|_{min}\ .
\eeq
Already at this point we learn from 
(\ref{bosonictrace}) and (\ref{fermionictrace})
 that 
for $D^a|_{min} =0$ 
we have a sum rule
\beq
\sum_{bosons} M_b^2\ =\ 2 \sum_{fermions} M_f^2 
\eeq
where in the sum real bosons are counted.

For $D^a|_{min} \neq 0$ also the gauge symmetry 
is necessarily broken and some of the gauge 
bosons become massive.
{}From  (\ref{covD})-(\ref{cvlagrphys})
one obtains
the mass matrix of the gauge bosons
\beq\label{vectormass}
M_{1}^2 = 2 g^2 \bA^j T^a_{jl} T^b_{lk} A^k
= 2 D^a_l D^b_{\bar l}
\eeq
Combining (\ref{bosonictrace}), 
(\ref{fermionictrace})
and (\ref{vectormass}) 
one arrives at the  mass sum rule \cite{Msf}:
\beq \label{str}
{\rm Str} M^2 \equiv \sum_{J=0}^1 (-)^{2J} (2J+1) \tr M_J^2 = -2 g (\tr T^a) D^a\ , 
\eeq
where $J$ is the spin of the particles.
The right hand side of (\ref{str})
vanishes for any non-Abelian factor 
in the gauge group 
while for $U(1)$ factors
it is proportional to the 
sum of the $U(1)$ charges $\sum Q_{U(1)}$.
Whenever this sum is non-vanishing
the theory has a $U(1)$ trace-anomaly.
(In the supersymmetric Standard Model this 
trace-anomaly vanishes.)
Finally, by repeating the steps of this section
one can show that (\ref{str}) holds over all field space
and not only at the minimum of $V$. This will play 
a r\^ole in deriving the soft supersymmetry breaking terms.
\bigskip

\noindent
{\bf Exercise:} 
{\it Verify (\ref{vectormass}) and 
(\ref{str}).} \\
{\bf Exercise:} 
{\it Compute $\sum Q_{U(1)_Y}$ in the supersymmetric
Standard Model.}\\
{\bf Exercise:} 
{\it Show that (\ref{str}) holds over all field space
and not only at the minimum of $V$.}

\bigskip

The sum rule (\ref{str})
is problematic for the supersymmetric Standard Model.
Since non of the supersymmetric partners 
has been observed yet they must be heavier
than the particles of the  Standard Model. 
Close inspection of (\ref{str}) shows that
this cannot be arranged within a spontaneously
broken supersymmetric Standard Model.

An additional problem is the presence of
 a massless Goldstone fermion.
Goldstone's theorem implies that any spontaneously
broken global symmetry leads to a massless state 
in the spectrum. This also holds for
supersymmetry where the broken generator is a Weyl spinor and thus there is an
 additional massless Goldstone fermion.
The presence of this state can be seen
explicitly from the condition that at the minimum
of the potential one has
\beq\label{gold}
V_j = W_{ij}  \bW_{{i}} + D^a_j D^a = 0 \ .
\eeq
Let us consider for simplicity the case
that supersymmetry is broken by a non-vanishing
F-term $\langle F_i\rangle = - \bW_i|_{\rm min} \neq 0$
while $\langle D^a\rangle =0$.\footnote{The general
case is discussed in ref.\ \cite{Msf}.}
{}From (\ref{gold}) one learns immediately
that now $ W_{ij}|_{\rm min}$ has to have a zero
eigenvalue. Using (\ref{fermionictrace})
this implies that also the mass matrix of 
the fermions has to have a zero
eigenvalue which is the Goldstone fermion. 

To summarize, the lesson of this section
is that also spontaneously broken supersymmetry runs
into phenomenological difficulties.
The only way out is an explicit breaking
of (global) supersymmetry.

\section{Extensions of the 
Standard Model with Softly Broken Supersymmetry}
\subsection{The Hierarchy and Naturalness Problem} \label{hn1}

Before we continue in our endeavor to construct
a phenomenologically viable extension
of the Standard Model let us briefly review 
what is called the hierarchy and 
naturalness problem in the 
Standard Model.\footnote{The discussion of this section
follows ref.\ \cite{RBg}.}

\begin{figure}[t]
\hspace*{1.1truein}
\psfig{figure=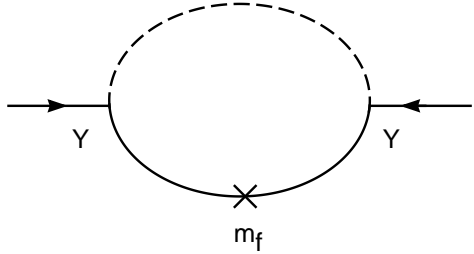,height=1.5in}
\caption{The one-loop correction to the fermion mass.}
\label{Fmass}
\end{figure}

Consider 
the following (non-supersymmetric)
Lagrangian of a complex scalar $A$
and a Weyl fermion $\chi$
\begin{eqnarray}
\cL  = & - & \partial_m \bA \partial^m A -i \bchi \bs^m
\partial_m \chi - \frac{1}{2} \,m_f\, (\chi\chi + \bchi\bchi)- 
 m^2_b\, \bA A \nonumber \\
& - & \;Y\,(A\chi\chi + \bA \bchi\bchi) \;-\; \lambda \, (\bA A)^2\ .
\label{toy model}
\end{eqnarray}
{}From (\ref{Lafter}) we learn that
this Lagrangian is supersymmetric if  $m_f=m_b$ and $Y^2=\lambda$ but let us not consider
this choice of parameters at first.
$\cL$  has a chiral symmetry for $m_f=0$ given by
\beq
A  \rightarrow  e^{-2 i \alpha}\, A\ , \qquad
\chi  \rightarrow e^{i \alpha}\, \chi \ .
\eeq
This symmetry prohibits the generation
of a fermion mass by quantum corrections.
For $m_f\neq 0$ the fermion 
mass does receive radiative corrections, but all possible diagrams  
have to contain a mass insertion as can be seen
from the one-loop diagram  shown 
in Fig.~\ref{Fmass}. Since the propagator of
the boson (upper dashed line  in the diagram)
is $\sim \frac{1}{k^2} $ while 
the propagator of the fermion (lower solid line) is
$\sim \frac{1}{k} $ one obtains a 
mass correction
which is proportional to $m_f$
\beq
\delta m_f \sim Y^2 m_f \ln {m_f^2\over \Lambda^2}\ ,
\eeq
where  $\Lambda$ is the ultraviolet cutoff.
Hence the mass of a chiral  fermion 
does not receive large radiative corrections
if the bare mass is small. 
For that reason `t~Hooft calls
fermion masses ``natural'' -- 
an extra symmetry appears
when the mass is set to zero
which in turn leads to a
protection of the fermion mass 
by an approximate
chiral symmetry \cite{thooft}.

\begin{figure}[t]
\hspace*{0.8truein}
\psfig{figure=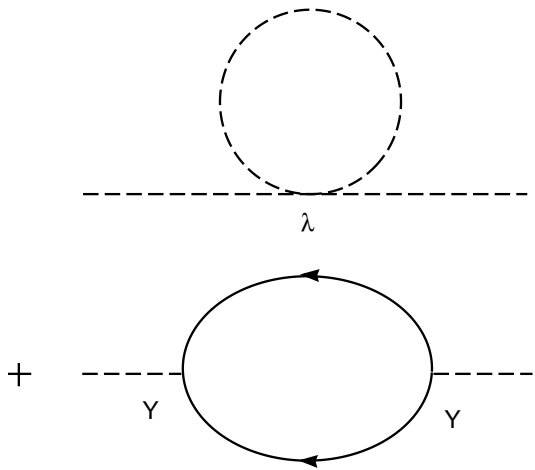,height=2.5in}
\caption{The one-loop corrections to the boson mass.}
\label{Bmass}
\end{figure}

This state of affairs is different for scalar fields. 
The diagrams giving the
one-loop corrections to $m_b$
are shown in Fig.~\ref{Bmass}.
Both diagrams are quadratically divergent but
they have an opposite sign
because in the second diagram fermions 
are running in the loop. One finds 
\beq \label{bmasscorrection}
\delta m_b^2 \sim  (\lambda  -  ~Y^2)\, \Lambda^2\ .
\eeq
Thus, in non-supersymmetric theories
scalar fields receive 
large mass corrections
(even if the bare mass is set to zero)
and small scalar masses are
``unnatural'' \cite{thooft,hier,tech}.
They can only be arranged  by delicately
fine-tuning the bare mass and the 
couplings $\lambda, Y$.
This problem becomes apparent in extensions
of the Standard Model which apart from 
the weak scale $M_Z$
do have a second larger scale, say 
$M_{\rm GUT}$ with $M_{\rm GUT}\gg M_{Z}$ 
\cite{hier,tech}.
In such theories the mass of the scalar boson
is naturally of the order of the largest
mass parameter in the theory.
This discussion applies 
to the Higgs boson of the Standard Model and
 it is difficult to understand the smallness
of $M_Z$ and how it can be kept stable
against quantum corrections
whenever the Standard Model
is the low energy limit of a theory
with a large mass scale.

A concrete example of this problem 
occurs in Grand Unified Theories (GUTs) \cite{ross}
where the Standard Model is embedded into a single
simple gauge group $G_{\rm GUT}$
(eg.\ $G_{\rm GUT}= SU(5)$).
The GUT gauge symmetry is  broken by 
a  Higgs mechanism to the
gauge group of the Standard Model and
 one has the following pattern of symmetry
breaking
\beq
G_{\rm GUT} \stackrel{M_{\rm GUT}}{\rightarrow} SU(3) \times SU(2) \times U(1) \stackrel{M_{Z}}{\rightarrow}
SU(3) \times U(1)_{\rm em}\ ,
\eeq
where $M_{\rm GUT} \approx 10^{15}$ GeV 
and thus $M_{\rm GUT} \gg M_Z$. 

There are basically two different
suggestions for `solving' this problem. 
The first class of models 
assume that the Higgs boson of the Standard Model is not 
an elementary scalar, but
rather a condensate of strongly interacting
`techni'- fermions \cite{tech,FS}. 
These theories are called  ``technicolor'' theories
but in all such theories it is difficult to
arrange agreement with the electroweak
precision measurements
of this decade \cite{RPes}.
The second class of models are supersymmetric
theories where the Higgs boson is elementary
but the quadratic divergence in 
(\ref{bmasscorrection})  exactly cancels 
due to the supersymmetric relation
$Y^2= \lambda$.

{\setcounter{footnote}{0}}
The cancellation of quadratic divergences 
is a general
feature of supersymmetric quantum field theories
and a  consequence of a more general 
non-renorm\-alization theorem:
The superpotential $W$ of a supersymmetric 
quantum field theory is not renormalized
in perturbation theory \cite{gsr} and 
all quantum corrections solely
arise from the gauge coupling 
and wavefunction renormalization.\footnote{This
non-renormalization theorem 
only holds in perturbation theory
but non-perturbative corrections do appear
 \cite{ads}.}
The non-renormalization theorem 
or in other words  the `taming' of the quantum 
corrections
is one of the attractive features of
supersymmetric quantum field theories.
It leads (among other things) to the possibility
of stabelizing the weak scale $M_Z$.

In that sense supersymmetry solves the 
naturalness problem in that it allows
for a small and stable weak scale without
fine-tuning. However, supersymmetry 
does not solve the hierarchy problem 
in that it does not explain why the weak scale
is small in the first place.

\subsection{Soft Breaking of Supersymmetry} \label{soft} 
As we have seen in section \ref{spon}
models with spontaneously broken supersymmetry are phenomenologically 
not acceptable. For example the mass formula (\ref{str}), 
generally valid in such cases,
forbids that all supersymmetric particles acquire masses large enough to make 
them invisible in present experiments. One way to overcome those difficulties 
is to allow explicit supersymmetry breaking. 

In the last section we observed 
that the absence of quadratic divergences 
in supersymmetric theories
stabilizes the Higgs mass and 
thus the weak scale.
This `attractive' feature of supersymmetric
field theories can be maintained in 
theories with explicitly broken 
supersymmetry if the
supersymmetry breaking terms are 
of a particular form.
Such terms which
break supersymmetry explicitely
and generate no quadratic divergences are called 
`soft breaking terms'. 

One possibility to identify the soft breaking terms is to
investigate the
divergence structure of the effective potential \cite{GG}.
 Consider a 
quantum field theory of a scalar field $\phi$ in the presence of an 
external source $J$. The generating functional for the Green's 
functions is given by
\begin{equation}
           e^{-iE[J]}\ =\ \int{\cal D}\phi\,
\mbox{exp}\left[i\int d^4x
                         ({\cal L}[\phi(x)]+J(x)\phi(x))\right] \mbox{ }.
\end{equation}
The effective action $\Gamma(\phi_{cl})$ is defined
by the  Legendre transformation
\begin{equation}
               \Gamma(\phi_{cl})=-E[J]-\int d^4xJ(x)\phi_{cl}(x) \mbox{ },
\end{equation}
where  $\phi_{cl} = - \frac{\delta E[J]}{\delta J(x)}$.
$\Gamma(\phi_{cl})$ can be expanded
in powers of momentum; in position
space this expansion takes the form
\begin{equation}
       \Gamma(\phi_{cl})=\int d^4x[-V_{eff}(\phi_{cl})
-\frac{1}{2}(\partial_m
        \phi_{cl})(\partial^m\phi_{cl})Z(\phi_{cl})+\dots\mbox{ }] \mbox{ }.
\end{equation} 
The term without derivatives is called the effective 
potential  $V_{eff}(\phi_{cl})$. 
It can be calculated in a  perturbation theory 
of  $\hbar$:
\begin{equation}
     V_{eff}(\phi_{cl})=V^{(0)}(\phi_{cl})+\hbar  V^{(1)}(\phi_{cl}) + \dots
\end{equation}
where $V^{(0)}(\phi_{cl})$ is the tree level and $V^{(1)}(\phi_{cl})$ the
one-loop contribution.
In a theory with scalars, fermions  and vector bosons 
 the 
one-loop contribution  takes the form \cite{coleman}
\beq \label{A}
V^{(1)}\sim \int d^4k\,\mbox{Str}  \ln(k^2+M^2) 
   = \sum_J(-1)^{2J}(2J+1)\, \mbox{Tr}
                    \int d^4k\, \ln(k^2+M_J^2)
\eeq
where $M_J^2$ is the matrix of second derivatives
of ${\cal L}|_{k=0}$ at zero momentum for scalars ($J=0$),
fermions ($J=1/2$) and vector bosons 
($J=1$).\footnote{$M_J^2$ is not necessarily
evaluated at the
minimum of $V_{eff}$. Rather it is a function of 
the scalar fields in the theory. The mass matrix
is obtained from $M_J^2$ 
by inserting the vacuum expectation 
values of the scalar fields.}
The UV divergences of (\ref{A}) can be displayed
by  expanding  the integrand 
in powers of large $k$. This leads to
\beq \label{B}
     V^{(1)} \sim \mbox{Str}{\bf 1} \int\frac{d^4k}{(2\pi)^4}\ln k^2
           + \mbox{Str}M^2\int\frac{d^4k}{(2\pi)^4} k^{-2} + \ldots\ .
\eeq
If a UV-cutoff $\Lambda$ is introduced 
the first term  in (\ref{B}) is 
${\cal O}(\Lambda^4\ln \Lambda)$.
Its coefficient Str${\bf 1}=n_B-n_F$
vanishes in theories with  a supersymmetric spectrum
of particles (cf.\ (\ref{nBF})). 
The second term in (\ref{B}) is ${\cal O}(\Lambda^2)$ and 
determines the presence of quadratic divergences at one-loop level.
Therefore quadratic divergences are absent 
if
\begin{equation} \label{C}\mbox{Str}M^2=0\ .
 \end{equation} 
More precisely, one can also
tolerate $\mbox{Str}M^2= const.$~ since
this would correspond to a shift of the zero
point energy which without coupling to gravity
is undetermined.
In theories with exact or spontaneously 
broken supersymmetry (\ref{C}) is 
fulfilled whenever the trace-anomaly vanishes
as we learned in (\ref{str}).\footnote{Indeed, 
theories with a non-vanishing D-term
have been shown to produce a quadratic 
divergence at one-loop
\cite{polchinski}. } 

The soft supersymmetry breaking terms are
defined as those non-supersym\-metric
terms that can be added to a supersymmetric
Lagrangian
without spoiling $\mbox{Str}M^2= const.$~.
One finds the following possibilities \cite{GG}
\begin{itemize}
\item
Holomorphic terms of the scalars
proportional to $A^2$, $A^3$ 
and the 
corresponding complex conjugates.\footnote{Higher 
powers of $A$ are forbidden since they generate 
quadratic divergences at the 2-loop level \cite{GG}.}
\item
Mass terms for the scalars proportional 
to $\bar{A} A$.\\
(They only contribute a constant, field independent
piece in $\mbox{Str}M^2$).
\item
Gaugino mass terms.\\
(A generic mass matrix of the fermions takes the form
\beq\label{dmass}
M_{1/2}=\left(\begin{array}{cc}W_{ij} + \delta W_{ij} & i\sqrt{2}D_i^b
   +\delta D_i^b \\   
 i\sqrt{2}D_j^a+\delta D_j^a & \delta\tilde m_{ab} \end{array} 
\right)  \mbox{ }, 
\eeq
where according to (\ref{fermmass})
\[ \left(\begin{array}{cc}W_{ij} & i\sqrt{2}D_i^b \\   
 i\sqrt{2}D_j^a & 0 \end{array} \right)  \mbox{ } \]
is the supersymmetric part of $M_{1/2}$.
Computing the supertrace of  (\ref{dmass})
reveals  that $\mbox{Str}M^2 = const.$
requires $\delta W=0=\delta D$ while
$\delta \tilde m$ can be arbitrary.)

\end{itemize}

Thus the most general Lagrangian with softly
broken supersymmetry  takes the form
\begin{equation} \label{Lsusy}
{\cal L}={\cal L}_{\rm susy}+{\cal L}_{\rm soft} \mbox{ }, 
\end{equation}
where ${\cal L}_{\rm susy}$ is of the form
(\ref{cvlagrphys}) and 
\begin{eqnarray} \label{Lsoft}
 {\cal L}_{\rm soft}&=&
- m^2_{ij}A^i\bar{A}^j - 
(b_{ij}A^iA^j + a_{ijk}A^iA^jA^k+\mbox{ h.c. })
 \nonumber\\                    
  &&   {}-\frac{1}{2}\, \tilde{m}_{ab}\lambda^a\lambda^b +\mbox{ h.c. }\mbox{ }.
\end{eqnarray} 
$m^2_{ij}$ and $b_{ij}$ are mass matrices for the scalars, $a_{ijk}$ are 
trilinear couplings (often called `A-terms')
and $\tilde{m}_{ab}$ is a mass 
matrix for the gauginos.

The next step will be to investigate 
if the more general Lagrangian
(\ref{Lsusy}) can be used to construct viable phenomenology.
Before we do so let us mention
that there is an alternative 
way to motivate the relevance of 
softly broken supersymmetric theories.
Ultimately one has to couple the supersymmetric
Standard Model to gravity.
This requires the  promotion of global
supersymmetry  to a local symmetry, that is the 
parameter of the supersymmetry transformation 
$\xi_{\alpha}=\xi_{\alpha}(x)$ 
is no longer constant but depends on 
the space-time coordinates $x$ \cite{IN,wb}. 
This demands the presence of an additional
 massless
fermionic gauge field (the gravitino)
$\Psi_{m\alpha}$ with spin 3/2
and an inhomogeneous  transformation law 
\beq
\delta_{\xi}\Psi_{m\alpha}=-\partial_m\xi_{\alpha}+\ldots\ .
\eeq
(The necessity of this transformation law can be 
seen for example from the supersymmetry
transformation of $\partial_m A$
which now has an extra contribution
$\partial_m\delta_{\xi}A\propto\partial_m\xi\psi=\xi\partial_m\psi
          +(\partial_m\xi)\psi \mbox{ }$.)    
Together with the metric  $g_{mn}$ 
and 6 auxiliary fields $b_m,M,\bar M$
the gravitino $\Psi_{m\alpha}$  
forms the supergravity multiplet
$(g_{mn},\Psi_{m\alpha},b_m,M,\bar M) \mbox{ }$.

The potential for the scalar fields 
is modified in the presence of supergravity
and found to be \cite{cremmer}
\begin{equation}
V(A,\bar{A})=e^{\kappa^2A\bar{A}}\Big[(D_iW)(\bar{D}_{\bar i}\bar{W})
            -3\kappa^2|W|^2\Big] 
+\frac12 D^a D^a\ ,
\end{equation}
where
\beq 
\kappa^2=\frac{8\pi}{M_{Pl}^2}\ , \qquad
D_iW=\frac{\partial W}{\partial A^i}+\kappa^2
\bar A^iW\ .
\eeq
The limit $\kappa^2\rightarrow 0$
corresponds to turning off gravity
and in this limit one obtains indeed 
$V\rightarrow  \frac{\partial W}{\partial A^i} 
\frac{\partial \bar{W}}{\partial \bar{A^i}}
+\frac12 D^a D^a$ 
in accord with (\ref{scalarpot}).
Local supersymmetry 
is spontaneously broken if
$D_i W|_{min}\not=0$
for some $i$.  
This can be achieved by introducing 
a hidden sector which only couples
via non-renormalizable interactions
to the observable sector
of the supersymmetric Standard Model 
and which has a superpotential
$W_{hid}(\phi)$ suitably chosen
to ensure 
$D_\phi W|_{min}\not=0$ \cite{HLW,RNi}.
In this case the 
gravitino becomes massive through 
a supersymmetric Higgs effect \cite{cremmer}.

In the limit
$\kappa^2\rightarrow 0$
with the gravitino mass  $m_{3/2}$ kept fixed
the Lagrangian for the fields
in the observable sector
looks precisely like eqs.\
(\ref{Lsusy}), (\ref{Lsoft})  \cite{HLW,RNi}.
 Thus the spontaneous breakdown of supergravity in a hidden sector 
manifests itself as explicit but soft breakdown of global supersymmetry in 
the low energy limit of the observable sector.

Finally, a variant of this mechanism
is to break supersymmetry dynamically
(ie.\ non-perturbatively) in an additional gauge sector
with some asymptotically free gauge theory \cite{ads,dns}.
In this case the supersymmetry breaking is communicated
to the observable sector by renormalizable interactions
but as in the previous case the breaking appears
in the observable sector as explicit but soft \cite{RDn,Rdynamical}.

\subsection{The 
Supersymmetric Standard Model with Softly Broken
Supersymmetry}
In the previous section we recalled 
the most general
Lagrangian of a softly broken supersymmetric
gauge theory in eqs.~(\ref{Lsusy}) and 
(\ref{Lsoft}). 
For ${\cal L}_{\rm susy}$ we continue to
take (\ref{cvlagrphys}) together with
the superpotential specified in
(\ref{mssmpot}).
For  ${\cal L}_{\rm soft}$ 
only  gauge invariance and R-parity is imposed.
This leads to the following
possible soft terms \cite{RNi,RHK,RHa,RZw,RBg,RDn}
\begin{eqnarray} \label{softSM} 
{\cal L}_{\rm soft}&=&{}
-\left((a_u)_{IJ}h_u\tilde{q}_L^I\tilde{u}_R^J
+(a_d)_{IJ}h_d\tilde{q}_L^I\tilde{d}_R^J
+(a_e)_{IJ}h_d \tilde{l}_L^I\tilde{e}_R^J
+ b h_uh_d + \mbox{ h.c. }\right) \nonumber\\
     &&{} -\sum_{\mbox{\footnotesize all scalars}} m_{ij}^2A^i\bar{A}^j-
         (\frac{1}{2}\sum_{(a)=1}^3
                 \tilde{m}_{(a)}(\lambda\lambda)_{(a)}+\mbox{ h.c. }) \mbox{ }.
\end{eqnarray}
Obviously a huge number of new
parameters is introduced via
${\cal L}_{\rm soft}$.
The parameters of ${\cal L}_{\rm susy}$
are the Yukawa couplings $Y$ and the 
parameter $\mu$ in the Higgs potential.
The Yukawa couplings are determined experimentally
already in the non-supersymmetric
Standard Model. In the softly broken
supersymmetric Standard Model
the parameter space is enlarged
by
\beq\label{softpara}
\left(\mu,(a_u)_{IJ}, (a_d)_{IJ}, (a_e)_{IJ}, b,
m^2_{ij}, \tilde{m}_{(a)}\right) \ .
\eeq
Not all of these parameters can be arbitrary
but quite a number of them
are experimentally constrained.
Some of these constraints we will see 
in the following sections.

Within this much larger parameter space
it is possible to overcome  several of the 
problems encountered in the supersymmetric Standard Model.
For example, the
supersymmetric particles can now easily be heavy
(due to the arbitrariness of the
mass terms $m^2_{ij}$)
and therefore out of reach of present experiments. 
Furthermore, the Higgs potential is changed
and vacua with spontaneous
 electroweak symmetry breaking can be arranged.

However, the soft breaking terms introduce their own set of difficulties.
For generic values of the parameters 
(\ref{softpara}) the contribution
to  flavor-changing neutral currents 
is unacceptably large \cite{ElN,RFCNCP}, 
additional (and forbidden)
sources of CP-violation occur \cite{EFN,RCP} and 
finally the absence of vacua
which break the $U(1)_{\rm em}$ and/or $SU(3)$
is no longer automatic \cite{Color}.
It is beyond the scope of these lectures
to review all of these aspects in detail.
Let us therefore focus on a few selected
topics and refer the reader to the literature
for further details and discussions.

\subsection{Electroweak Symmetry Breaking}
In section~3.2 we noticed that for unbroken
or spontaneously broken supersymmetry
the electroweak symmetry remains intact in
the supersymmetric version of the Standard Model.
Let us now review the situation in the presence
of soft breaking terms \cite{HHG}.
The Higgs sector of the supersymmetric
Standard Model consists of two $SU(2)$-doublets
\[ h_u={h_u^+\choose h_u^0} \quad,\quad\quad h_d={h_d^0\choose h_d^-} \quad, \]
which carry eight real degrees of freedom, four of them neutral and four 
charged. Like in the Standard Model $SU(2)_L\times U(1)_Y$ 
will be broken to $U(1)_{\rm em}$ 
by non-vanishing VEVs of the neutral Higgs bosons $h_u^0$ and $h_d^0$. 
For that purpose consider their potential
which can be derived from eqs.~(\ref{HiggsV}),
(\ref{softSM})
\begin{eqnarray} \label{higgspot}
V(h_u^0,h_d^0)= \hat{m}_u^2|h_u^0|^2
+\hat{m}_d^2|h_d^0|^2
-b(h_u^0h_d^0+\bar{h}_u^0\bar{h}_d^0) 
+\frac{g_1^2+g_2^2}{8}
 \Big(|h_u^0|^2-|h_d^0|^2\Big)^2,
\end{eqnarray} 
where
\begin{eqnarray} 
\hat{m}_u^2&=&m_u^2+|\mu|^2 \mbox{ },\nonumber\\
                            \hat{m}_d^2&=&m_d^2+|\mu|^2 \mbox{ }.
\end{eqnarray}
Notice that the coefficient of the $|h_{u,d}^0|^4$-term is exactly as in
(\ref{HiggsV})  determined by the gauge couplings
and not changed by soft breaking terms.
This term is positive 
so that the potential is bounded from below 
for large values of  $h_u^0,h_d^0$ as long as
$|h_u^0|\not=|h_d^0|$. 
To secure this bound also in the direction
$|h_u^0|=|h_d^0|$ one has to impose the
following constraint on the parameter space
\begin{equation}  \label{bb}
\hat{m}_u^2+\hat{m}_d^2 \ge 2|b| \mbox{ }.
\end{equation} 

\bigskip

\noindent
{\bf Exercise:} 
{\it  Verify formula (\ref{higgspot}) using  
eqs.~(\ref{HiggsV}),
(\ref{softSM}).}\\
{\bf Exercise:} 
{\it Verify the condition (\ref{bb}).}
\bigskip

The existence of a minimum
of $V$ with broken gauge symmetry
requires that the
Hessian of $V(h_u^0,h_d^0)|_{h_u^0=h_d^0=0}$ 
\beq
 \left( \begin{array}{cc} \hat{m}_u^2 & -b \\ -b & \hat{m}_d^2 
\end{array}\right)
\eeq
has at least one negative eigenvalue. 
Together with  (\ref{bb}) 
this implies\footnote{This is the generalization
of the condition $\mu^2 <0$ in the Standard Model
to a Higgs sector with two Higgs doublets.}  
\begin{equation} \label{sb}\hat{m}_u^2\hat{m}_d^2<b^2 \mbox{ }.
\end{equation}
So the soft terms have to satisfy 
(\ref{bb}), (\ref{sb}) in order to  
induce electroweak symmetry breaking
but in addition also the 
masses of the $Z$- and $W$-bosons
have to come out correctly.
These masses are given by
\begin{eqnarray} \label{D}
M_Z^2&=&\frac{1}{4}(g_1^2+g_2^2)(v_u^2+v_d^2)=\frac{1}{2}(g_1^2+g_2^2)v^2
\nonumber \\
M_W^2&=&\frac{1}{4}g_2^2(v_u^2+v_d^2)=\frac{1}{2}g_2^2v^2 \mbox{ },
\end{eqnarray}
where 
\begin{equation} \label{E}
\langle h_u^0\rangle =\frac{1}{\sqrt{2}}\,v_u=v\sin\beta,\qquad 
 \langle h_d^0\rangle =\frac{1}{\sqrt{2}}\,v_d=v\cos\beta\ .
\end{equation}
The electroweak symmetry breaking is parameterized
by the two Higgs vacuum expectation values
$v_u,v_d$ (which can be chosen real)
or equivalently $v$ and $\beta$.
As in the Standard Model $v$ has to be chosen
such that it reproduces the experimentally
measured $Z$- and $W$-masses.
$\beta$ on the other hand is an additional parameter
which arises as a consequence of the enlarged
(2 doublet) Higgs sector.

The Higgs expectation values are directly
related to the parameters of the Higgs potential
via the minimization conditions
\begin{eqnarray}\label{F}
\frac{\partial V}{\partial h_u^0}&=&2\hat{m}_u^2v_u-2bv_d
+\frac{g_1^2+g_2^2}{2}(v_u^2-v_d^2)v_u=0\ ,
 \nonumber\\
\frac{\partial V}{\partial h_d^0}&=&2\hat{m}_d^2v_d-2bv_u
-\frac{g_1^2+g_2^2}{2}(v_u^2-v_d^2)v_d=0 \mbox{ }.
\end{eqnarray}
This in turn can be used to derive a more physical
relationship among the parameters.
Using (\ref{D}) and (\ref{E}) the minimization conditions (\ref{F}) can
be rewritten as
\begin{eqnarray} \label{MZcond}
b&=&\frac{1}{2}\sin 2\beta\, (\hat{m}_u^2+\hat{m}_d^2)\nonumber \\ \label{H}
M_Z^2&=&-2|\mu|^2+\frac{2}{1-\tan^2\beta}(m_u^2\tan^2\beta-m_d^2) \mbox{ }.
\end{eqnarray}

Finally, the full Higgs potential including
all eight real degrees of freedom
can be used to compute the $8\times8$
mass matrix of all Higgs bosons.
After a somewhat lengthy calculation \cite{HHG}
one finds that this mass matrix has 
three eigenvalues zero corresponding
to the three Goldstone
modes `eaten' by the $W^{\pm}$ and the $Z$.
The remaining five degrees
of freedom yield the physical Higgs bosons of the 
model:
\begin{eqnarray}
H^{\pm}&&\quad\quad\mbox{charged Higgs boson pair}\nonumber \\
A^0&&\quad\quad\mbox{CP-odd neutral Higgs boson}\nonumber \\
H^0,h^0&&\quad\quad\mbox{CP-even neutral Higgs bosons .}\nonumber
\end{eqnarray}
Their  tree-level masses are given by
\begin{eqnarray} \label{G}
m_A^2&=&\hat{m}_u^2+\hat{m}_d^2 \nonumber\\
m_{H_{\pm}}^2&=&m_A^2+M_W^2 \nonumber\\
m_{h^0}^2&=&\frac{1}{2}\Big[m_A^2+M_Z^2-\sqrt{(m_A^2+M_Z^2)^2
-4m_A^2M_Z^2\cos^22\beta}\mbox{ }\Big] \nonumber\\
m_{H^0}^2&=&\frac{1}{2}\Big[m_A^2+M_Z^2+\sqrt{(m_A^2+M_Z^2)^2
-4m_A^2M_Z^2\cos^22\beta}\mbox{ }\Big]\mbox{ }.
\end{eqnarray}
The Higgs masses obey physically interesting
mass relations.
{}From (\ref{G}) we learn 
\begin{equation}\label{massrel}
m_{H^{\pm}}\ge M_W\ , \qquad
m_{H^0}\ge M_Z\ , \qquad
m_{h^0} \le M_Z \ .
\end{equation}

\bigskip
\noindent
{\bf Exercise:} {\it Derive the mass relations
(\ref{massrel}) from (\ref{G}).}
\bigskip

Physically the most interesting is the last inequality since it predicts the
existence of a light Higgs boson.
This `prediction' can be directly traced to 
the fact the quartic couplings in the Higgs potential
are fixed by the (measured) gauge couplings and 
are not free parameters as in the Standard Model.
However, radiative corrections 
for this lightest Higgs boson mass can be large
 and after taking into
account quantum corrections
the upper bound on $m_{h^0}$ is pushed up
to about 150 GeV \cite{hrad}.
However, the prediction of one light neutral Higgs
boson remains and is one of the characteristic
features of the supersymmetric
two-doublet Higgs sector.
It even holds in the limit that all 
masses of the supersymmetric particles
are sent to infinity.
In this limit one recovers the non-supersymmetric
Standard Model -- albeit with a light Higgs. 

Finally, a proper minimization of the scalar
potential requires to take into account
all scalar fields 
and not truncate to the neutral Higgs
bosons. It is possible (and does occur
in certain regions of the 
supersymmetric parameter space)
that there exist minima which not only break
the electroweak symmetry but also
$SU(3)$ and/or $U(1)_{\rm em}$.
An extended analysis of this aspect can be found, 
for example, in ref.\ \cite{Color}.

\subsection{Weak-Scale Supersymmetry}
Let us briefly come back to the hierarchy and 
naturalness problem. 
In section~\ref{hn1} we learned that
supersymmetry sheds no light 
on the hierarchy problem, i.e.\ the
question  why 
$M_Z$ is so much
smaller than $M_{Pl}$.
However, because
of the absence of quadratic divergences 
it does solve the naturalness problem,
that is it provides a stable 
hierarchy in the presence of a light Higgs boson.
To only add soft supersymmetry breaking terms
into the supersymmetric Standard Model
was precisely motivated by this feature.
However, from eq.~(\ref{H}) we learn 
an important additional constraint
on the soft breaking parameters.
In order to introduce no new fine-tuning
all soft terms in  eq.~(\ref{H}) should be 
of the same order of magnitude,
i.e.\ of order ${\cal O}(M_Z)$ or at most 
in the TeV range \cite{Wi}.
If this were not the case the soft parameters 
would have to be delicately tuned
in order to add up 
to $M_Z$. This in turn implies 
the breaking of supersymmetry should occur
at the weak scale and that most likely
all supersymmetric particles have masses 
in that range.\footnote{This argument is
somewhat imprecise since not only 
the soft breaking terms $(m_u,m_d)$
but also $\mu$ have to be  ${\cal O}(M_Z)$.
However $\mu$ is not related to supersymmetry
breaking in any obvious way but
rather a parameter in the superpotential.
Thus one needs
a mechanism which also 
explains the approximate equality
of $\mu$ with $(m_u,m_d)$.
This is known as the $\mu$-problem \cite{KN,RNi}.}

This line of argument is used
to motivate what is called
``weak-scale supersymmetry'' and indeed
the current LEP II experiments actively
look
for supersymmetric particles with masses
slightly above the weak scale.

\subsection{Further Constraints on the Supersymmetric Parameter Space}
The experimental searches for supersymmetric particles
impose additional constraints on the supersymmetric 
parameter space. First and foremost
the direct lower
bounds on the masses of the supersymmetric
particles \cite{PDG} 
exclude certain regions of the parameter space.
The translations of experimental bounds
into the supersymmetric parameter space is
complicated by the fact that 
the states which are listed in table~2 are
interaction eigenstates, but not necessarily mass eigenstates. 
The only exception are the gluinos $\tilde G$ and the mass bounds
directly translate into bounds on $\tilde m_3$.
On the other hand the three Winos $\tilde W^\pm, 
\tilde W^3$, the $\tilde B$ and the four 
Higgsinos  $\tilde h_{u,d}^0, \tilde h_{d}^-,
\tilde h_{u}^+$ combine into 
a four-vector of neutral Weyl fermions
consisting of
$ {\bf N} \equiv
(\tilde B, \tilde W^3,\tilde h_{u}^0,\tilde h_{d}^0)$
and two pairs of charged Weyl fermions
${\bf C}^- \equiv
(\tilde W^-, \tilde h_{d}^-),\ 
{\bf C}^+ \equiv (\tilde W^+,\tilde h_{u}^+)$
with the following set of mass matrices
\beq
{\cal L}_{\rm fmass} =
-\frac12\, \tilde m_3\, \tilde G^a\tilde G^a \
-\ {\bf C}^- M_C ({\bf C}^+)^T
\ -\ \frac12\,  {\bf N}  M_N {\bf N}^T
 + h.c. \ ,
\eeq
where
\beq\label{mchargino}
M_C= \left(\begin{array}{cc}
\tilde m_2 & i\sqrt 2 g_2 v_u\\
i\sqrt 2 g_2 v_d& \mu\end{array}\right)\ ,
\eeq
and 
\beq\label{mneutralino}
M_N= \left(\begin{array}{cccc}
\tilde m_1 & 0&\frac{i}2 g_1 v_u& -\frac{i}2 g_1 v_d   \\
0&\tilde m_2& -\frac{i}2 g_2 v_u& \frac{i}2 g_2 v_d   \\
\frac{i}2 g_1 v_u& -\frac{i}2 g_2 v_u&0&\mu   \\
-\frac{i}2 g_1 v_d& \frac{i}2 g_2 v_d&\mu &0  
\end{array}\right)\ .
\eeq
Thus, the physical
mass eigenstates of $M_C$ and $M_N$
are parameter dependent linear 
combinations of the corresponding
interaction eigenstates and 
they are termed {\em charginos}
and {\em neutralinos}, respectively.

\bigskip
\noindent
{\bf Exercise:} {\it Derive the mass matrices
(\ref{mchargino}) and (\ref{mneutralino}).
}
\bigskip

The other supersymmetric particles
are the spin-0
partners of the quarks and leptons,
the {\em squarks} $\tilde q_L^I, \tilde u_R^I,\tilde d_R^I$
and the {\em sleptons}
$\tilde l_L^I, \tilde e_R^I$.
Their mass eigenstates are derived from
the following
three $6\times6$ and one $3\times3$ mass matrices
\beq
{\bf U} M_U {\bf U}^\dagger\ ,\qquad
{\bf D} M_D {\bf D}^\dagger\ ,\qquad
{\bf E} M_E {\bf E}^\dagger\ ,\qquad
\tilde \nu M_\nu \tilde{\bar \nu}\ ,
\eeq
where
\beq
{\bf U} \equiv (\tilde u_L^I, \bar{\tilde u}_R^I)
\ ,\qquad
{\bf D} \equiv (\tilde d_L^I, \bar{\tilde d}_R^I)
\ ,\qquad
{\bf E} \equiv (\tilde e_L^I, \bar{\tilde e}_R^I)
\ .
\eeq
Explicit forms of these mass
matrices in terms of the soft parameters
can be found e.g.~in ref.~\cite{RHa}.
Constraints on the soft parameters
imposed by the experimental bounds
can be found in \cite{PDG,RHa,RBg,RDn,RGu,RPes,Rlep}.

\subsection{The Minimal Supersymmetric Standard  \\ Model (MSSM)}
The supersymmetric version of the Standard Model we discussed
so far has a huge parameter space and therefore very limited
predictive power. A much more constrained version (with less free parameters)
became known as the Minimal Supersymmetric Standard Model (MSSM)
which is the topic of this section.
The MSSM was motivated  by the success of 
Grand Unified Theories combined with a simple, flavor blind mechanism
of supersymmetry breaking in a hidden sector 
\cite{DG}.
Over the last 15 years this model went through a few alterations but today
it is a well defined model with a very particular
set of soft supersymmetry breaking terms which are flavor blind
and in some sense the minimal choice of free parameters \cite{RNi,RHK,RHa,RZw}.
One imposes that {\it all} scalar masses are the same
$m^2_{ij}=m_0^2\delta_{ij}$, all gaugino masses are the same
$\tilde{m}_1=\tilde{m}_2=\tilde{m}_3=\tilde{m}$,
all a-parameters are proportional to the Yukawa couplings
with the same universal proportionality constant
$a_0$
and finally that the b-parameter is of a specific form.
Altogether one has
\begin{eqnarray}\label{MSSMsoft}
m^2_{ij}&=&m_0^2\, \delta_{ij}\ , \qquad
\tilde{m}_1=\tilde{m}_2=\tilde{m}_3=\tilde{m}
\ ,\qquad b=b_0\, m_0\, \mu\ ,\nonumber\\
(a_u)_{IJ}&=&a_0\,  (Y_u)_{IJ}\ , \qquad  
 (a_d)_{IJ}=a_0\,  (Y_d)_{IJ}\ , \qquad
(a_l)_{IJ}=a_0\,  (Y_l)_{IJ} \ .
\end{eqnarray}
Thus, the parameter space of the MSSM is 
spanned by the 5 parameters
\[(m_0,\tilde{m},a_0,b_0,\mu)\mbox{ },\]
which are subject to one further 
non-trivial constraint (\ref{MZcond})
which ensures 
the observed electroweak symmetry breaking.

Of course the relations (\ref{MSSMsoft}) 
are meant to be 
tree level relations and they do enjoy
quantum corrections which are governed
by the appropriate renormalization group equations
\cite{APW,RGEa}. 
The quantum corrections destroy 
the universality of the soft parameters 
but the deviation from universality is 
small and so far in accord with all
measurements \cite{RBg,RGu,Poko,RFCNCP}.
In particular, the smallness of 
flavor-changing neutral currents is
`naturally explained' in the MSSM \cite{ElN,BBM}.
Thus even without an underlying GUT theory
this set of parameters seems to have some
phenomenological attraction.

However, one should also stress that supersymmetric
GUTs are an extremely viable possibility today.
Among other things the LEP precision experiments 
determined the gauge coupling constant $g_2$ 
very precisely.
In light of these measurements
one can ask to what extend the three gauge 
couplings
$g_1,g_2,g_3$ do unify at some 
high energy scale $M_{\rm GUT}$
given the experimental input at $M_Z$.
At one-loop order the energy dependence
of the gauge couplings is given by
\begin{equation} 
g_{(a)}^{-2}(M_Z)\ =\ g_{(a)}^{-2}(M_{\rm GUT})\,
+\,  \frac{b_{(a)}}{8\pi^2}\, 
\ln \big(\frac{M_{\rm GUT}}{M_Z}\big) \ ,
\end{equation}
where $b_{(a)}$ are the coefficients 
of the one-loop beta-function
which depend on the massless spectrum of the 
theory.
Let us first recall that the index $T_{(a)}(R)$ of
a representation $R$ is 
defined as $\mbox{Tr}_R(T^aT^b)_{(a)}=T_{(a)}(R)\delta^{ab}$ where $T^b$ are
the generators of the gauge group.
In terms of the indices $b_{(a)}$ is given by
\beq \label{b}
b_{(a)} =
-\frac{11}{3}T_{(a)}(\mbox{G})
+\frac{2}{3}T_{(a)}(\mbox{R})\, n_{WF}
+\frac{1}{6}T_{(a)}(\mbox{R})\, n_{RS} \mbox{ },
\eeq
where G denotes the adjoint
representation and
$n_{WF}$ ($n_{RS}$) counts the number of 
Weyl fermions (real scalars)
in the representation $R$.

For the non-supersymmetric Standard Model
one finds 
\[ (b_1,b_2,b_3)=(\frac{41}{10},-\frac{19}{6},-7) \mbox{ }, \]
which does not lead to a unification
of coupling constants at any scale. That is,
one cannot find an $M_{\rm GUT}$ where
$g_{(1)}(M_{\rm GUT})=g_{(2)}(M_{\rm GUT})=
g_{(3)}(M_{\rm GUT})$ holds.
However, in the supersymmetric Standard Model
one finds
\[ (b_1,b_2,b_3)=(11,1,-3) \mbox{ }, \]
which does lead to a unification of couplings
at $M_{GUT}\simeq 10^{16}$ GeV \cite{Uni}. 
This can be taken as a (strong) hint for a 
supersymmetric GUT.

Let us close this section with a discussion 
of electroweak symmetry breaking in the MSSM.
At the tree level 
one now has $\hat{m}_d^2=\hat{m}_u^2$ 
and as a consequence the conditions for
electroweak symmetry
breaking
(\ref{bb}) and (\ref{sb}) 
cannot be satisfied simultaneously 
\begin{eqnarray}\hat{m}_u^2+\hat{m}_d^2=2(m_0^2+\mu^2)\ge 2 |b| \nonumber\\
\hat{m}_u^2\hat{m}_d^2=(m_0^2+\mu^2)^2 < |b|^2 \mbox{ }.\nonumber
\end{eqnarray}
However, quantum corrections 
alter this situation and naturally induce 
electroweak symmetry breaking \cite{IR}.
Thus, the MSSM naturally displays 
 a supersymmetric version of the Coleman-Weinberg mechanism \cite{coleman}
where quantum corrections
generate a non-trivial minimum in the Higgs potential.
The electroweak symmetry breaking scale  
is  not
put in hand, but  related to the scale
of the supersymmetry breakdown.

\section{Summary}
Supersymmetry is a generalization of the space-time symmetries of quantum field theories
and it transforms fermions into bosons and 
vice versa. In the supersymmetric Standard Model
all particles of the Standard Model
are accompanied by superpartners 
with opposite statistics. 
Moreover, it is necessary to enlarge the 
Higgs sector and add a second Higgs doublet
to the spectrum. 
Supersymmetry cannot be an exact symmetry
of nature and has to be realized in its broken
phase (if at all).
However, spontaneously broken 
global supersymmetry is 
phenomenologically ruled out.
Spontaneously broken local supersymmetry on the other hand
leads to models which are (still) 
in agreement with all present observations. 
These models do provide
a solution to the naturalness problem as long as
the supersymmetric partners have masses
not much bigger than $1 TeV$. 
The supersymmetric Standard Model
has one solid prediction: a light neutral Higgs
with a mass smaller than 150 GeV
(as well as additional charged 
but not necessarily light Higgs bosons). 
The gauge couplings of 
the supersymmetric Standard Model do
unify at a scale $M_{GUT}\simeq 10^{16}$ GeV, 
which may indicate a supersymmetric GUT.
The MSSM -- motivated by GUTs --
contains only four new parameters
and is consistent with observations in large
regions of this parameter space. 
In this model the electroweak symmetry breaking 
is driven by radiative corrections realizing
a supersymmetric version of the Coleman-Weinberg
mechanism. 
Finally, the concept of supersymmetry also
 arises naturally in string theories
which might be another hint towards
its realization in nature.

\bigskip
\noindent
{\bf Exercise:} {\it  How can supersymmetry be verified or falsified? Distinguish between
necessary and sufficient conditions.
}
\bigskip

\newpage
\section{Appendix-Conventions and Notation}
In these lectures the notation and conventions 
of ref.\ \cite{wb} are used.
 The  four-dimensional Lorentz metric is chosen as
\begin{equation} \eta_{mn}=\mbox{diag}(-1,1,1,1) \ .\end{equation}
Lorentz indices are labeled by  Latin indices $m,n,...$ which run from 0 to 3.
Greek indices are used
to denote spinors. 
A two-component Weyl spinor can transform under the
$(\frac{1}{2},0)$ or the complex conjugate
$(0,\frac{1}{2})$--representation of the Lorentz group
and dotted or 
undotted indices are used to distinguish between these representations.
$\psi_{\alpha}$ denotes a spinor transforming under
the $(\frac{1}{2},0)$ representation while 
$\bar{\chi}_{\dot{\alpha}}$ 
transforms under the 
$(0,\frac{1}{2})$ representation of the Lorentz group.
The spinor indices $\alpha$ and $\dot{\alpha}$ can take the values 1 and 2.
These indices can be raised and lowered using the skew-symmetric
$SU(2)$-- invariant tensor
$\epsilon^{\alpha\beta}$ or  $\epsilon_{\alpha\beta}$.
\begin{equation} \psi^{\alpha}=\epsilon^{\alpha\beta}\psi_{\beta} \mbox{ }, \quad\quad 
                             \psi_{\alpha}=\epsilon_{\alpha\beta}\psi^{\beta} \mbox{ },
\end{equation}
where
$$\epsilon_{21}=-\epsilon_{12}=1,\quad
 \epsilon_{11}=\epsilon_{22}=0, \quad
\epsilon_{\alpha\gamma}\epsilon^{\gamma\beta}=\delta_{\alpha}^{\beta}\ .
$$
For dotted indices the analogous equations hold.
The product
$ \epsilon^{\beta \alpha} \psi_{\alpha} \chi_{\beta}= 
\psi^{\beta} \chi_{\beta}$ is a  Lorentz scalar.
Spinors are anticommuting objects and one has the
following summation convention:
\begin{eqnarray}&&\psi\chi=\psi^{\alpha}\chi_{\alpha}=-\psi_{\alpha}\chi^{\alpha}=\chi^{\alpha}\psi_{\alpha}=\chi\psi \mbox{ },\nonumber\\
                        &&  \bar{\psi}\bar{\chi}=\bar{\psi}_{\dot{\alpha}}\bar{\chi}^{\dot{\alpha}}=-\bar{\psi}^{\dot{\alpha}}\bar{\chi}_{\dot{\alpha}}=
                           \bar{\chi}_{\dot{\alpha}}\bar{\psi}^{\dot{\alpha}}=\bar{\chi}\bar{\psi} \mbox{ }.
\end{eqnarray}
 The convention for the conjugate spinors are
chosen such that it is consistent with the conjugation of scalars:
\begin{equation} (\psi\chi)^{\dagger}=(\psi^{\alpha}\chi_{\alpha})^{\dagger}=\bar{\psi}_{\dot{\alpha}}\bar{\chi}^{\dot{\alpha}}=
                                    \bar{\psi}\bar{\chi}=\bar{\chi}\bar{\psi}\mbox{ }.
\end{equation}

The $\sigma$-matrices $\sigma_{\alpha\dot{\alpha}}^m$ are given by:
\begin{eqnarray} \sigma^0=\left(\begin{array}{cc}-1&0 \\ 0&-1\end{array}\right)\mbox{ }, \quad\quad 
                            \sigma^1=\left(\begin{array}{cc}0&1 \\ 1&0\end{array}\right)\mbox{ }, \nonumber\\
                             \sigma^2=\left(\begin{array}{cc}0&-i \\ i&0\end{array}\right)\mbox{ }, \quad\quad 
                             \sigma^3=\left(\begin{array}{cc}1&0 \\ 0&-1\end{array}\right)\mbox{ }.
\end{eqnarray}
The invariant $\epsilon$--tensor raises and lowers their indices:
\begin{equation} \bar{\sigma}^{m\dot{\alpha}\alpha}=\epsilon^{\dot{\alpha}\dot{\beta}}\epsilon^{\alpha\beta}\sigma_{\beta\dot{\beta}}^m \mbox{ }
\end{equation}
and we have:
\beq
 \bar{\sigma}^0={\sigma}^0\ , \qquad                     \bar{\sigma}^{1,2,3}=-{\sigma}^{1,2,3} \mbox{ }.
\eeq
The generators of the Lorentz group in the spinor representation
are given by
\beq
\sigma^{nm} = \frac14
(\sigma^n\bar\sigma^m-\sigma^m\bar\sigma^n)\ , \qquad
\bar\sigma^{nm} = \frac14
(\bar\sigma^n\sigma^m-\bar\sigma^m\sigma^n)\ .
\eeq

The Dirac-$\gamma $--matrices can be written in terms of Weyl matrices:
\beq
\gamma^m = 
\left( \begin{array}{cc} 0 & \sigma^m \\ \bar{\sigma}^m & 0\end{array}
\right)
\eeq
which fulfill 
\beq
\{ \gamma^m, \gamma^n \} = -2 \, \eta^{mn}\ .
\eeq
A four component Dirac spinor contains two Weyl spinors 
\beq
\Psi_D = \left( \begin{array}{c} \psi \\ \bar{\chi} \end{array} \right)
= \left( \begin{array}{c} \psi_{\alpha} \\ \bar{\chi}^{\ad} \end{array} 
\right)\ .
\eeq
Its conjugate is
\beq
\bPsi_D = \Psi_D^{\dagger} \gamma^0 = 
(\chi^{\alpha},\bar{\psi}_{\ad})\ .
\eeq
The Dirac equation describing relativistic spin-$\frac{1}{2}$ particles reads:
\beq
(i \, \gamma^n \partial_{n} + m)\, \Psi_D =0.
\eeq
It can be decomposed into two Weyl equations
\bea
i \sigma^n \partial_n \, \bar{\chi} \, + \, m \psi &=& 0\ , \\ \nn
i \bar{\sigma^n} \partial_n \psi \, + \, m \bar{\chi} &=& 0\ .
\eea

\bigskip
\noindent
{\bf Exercise:} {\it
Show the validity of the following identities:}
 $$\psi^{\alpha} \chi_{\alpha} = \chi^{\alpha} \psi_{\alpha} , \quad
\chi^{\alpha} \sigma_{\alpha \ad}^n \bpsi^{\ad}
        = -\bpsi_{\ad} \bs^{n\ad \beta} \chi_{\beta},\quad
\left( \psi^{\alpha} \phi_{\alpha} \right) \bar{\chi}_{\bd}
        = -\frac{1}{2} \left( \phi^{\alpha} \sigma^m_{\alpha \dd} 
           \bar{\chi}^{\dd} \right) \psi^{\gamma} \sigma^m_{\gamma \bd}
$$

\noindent
{\bf Exercise:} {\it
Compute in terms of $\sigma$-matrices} the following
matrices
$$\gamma_5 = -i \gamma_0 \gamma_1 \gamma_2 \gamma_3 \qquad
 P_L= \frac{1}{2} (1-\gamma_5), \qquad P_R = \frac{1}{2} (1+\gamma_5)
$$
 
\noindent
{\bf Exercise:} {\it
 Express the following Lorentz-invariants in terms of Weyl-spinors:}
$$
\bPsi_D \Psi_D, \; \bPsi_D \gamma^5 \Psi_D, \; \bPsi_D \gamma^m \Psi_D, \;
\bPsi_D \gamma^5 \gamma^m \Psi_D, \; 
\bPsi_D [\gamma^m, \gamma^n] \Psi_D 
$$

\vskip 1cm

{\bf Acknowledgement:}
We would like to thank the students and organizers
of the 1996 Saalburg Summers School for 
generating an inspiring scientific atmosphere.
Special thanks go to H.\ G\"unther, 
M.\ Haack and  M.\ Klein 
for carefully reading the manuscript and 
P.\ Zerwas for guiding us through the 
CERN home page.
J.L.\ also thanks the students of the University of Munich
for their questions and comments during a course
on supersymmetry which was the seed of these lecture notes.

%

\end{document}